\documentclass[10pt,aps,prl,twocolumn,fleqn,superscriptaddress,nofootinbib,preprintnumbers,nobalancelastpage]{revtex4-1}
\pdfoutput=1
\usepackage{amsmath,amssymb,pstricks,graphicx,xspace,subfigure,ulem}

\newcommand{\MEPS}{M\scalebox{0.8}{E}P\scalebox{0.8}{S}@LO\xspace}
\newcommand{\MCatNLO}{M\protect\scalebox{0.8}{C}@N\protect\scalebox{0.8}{LO}\xspace}
\newcommand{\SMCatNLO}{S--M\protect\scalebox{0.8}{C}@N\protect\scalebox{0.8}{LO}\xspace}
\newcommand{\MEPSatNLO}{M\scalebox{0.8}{E}P\scalebox{0.8}{S}@NLO\xspace}

\newcommand{\Sherpa}{S\protect\scalebox{0.8}{HERPA}\xspace}
\newcommand{\SherpaOpenLoops}{S\protect\scalebox{0.8}{HERPA}+O\protect\scalebox{0.8}{PEN}L\protect\scalebox{0.8}{OOPS}\xspace}
\newcommand{\OpenLoops}{O\protect\scalebox{0.8}{PEN}L\protect\scalebox{0.8}{OOPS}\xspace}
\newcommand{\Comix}{C\protect\scalebox{0.8}{OMIX}\xspace}
\newcommand{\Amegic}{A\protect\scalebox{0.8}{MEGIC++}\xspace}
\newcommand{\Collier}{C\protect\scalebox{0.8}{OLLIER}\xspace}
\newcommand{\Rivet}{R\protect\scalebox{0.8}{IVET}\xspace}

\newcommand{\mc}[1]{\mathcal{#1}}

\newcommand{\ttbar}{t\bar{t}} 
\newcommand{\pp}{pp} 
\newcommand{\pT}{p_\mathrm{T}}
\newcommand{\resseperrs}[9]{$#1_{-#2-#3-#4-#5}^{+#6+#7+#8+#9}$ pb}
\newcommand{\Hl}{\vphantom{$\int\limits_a^b$}}
\usepackage[pdfborder={0 0 0}]{hyperref}
\hypersetup{
  pdfauthor={Stefan Hoeche, Frank Krauss, Philipp Maierhoefer, Stefano Pozzorini, Marek Schoenherr, Frank Siegert},
  pdftitle={Next-to-leading order QCD predictions for top-quark pair production
    with up to two jets merged with a parton shower},
  pdfkeywords={QCD, NLO, Parton Shower, Top Quark}
}

\begin{document}

\preprint{SLAC-PUB 15911}
\preprint{IPPP/14/18}
\preprint{DCPT/14/36}
\preprint{ZU-TH 06/14}
\preprint{LPN14-051}
\preprint{HT-14-005}
\preprint{MCNET-14-05}

\title{Next-to-leading order QCD predictions for top-quark pair production
       \texorpdfstring{\\}{}
       with up to two jets merged with a parton shower}

\author{Stefan~H{\"o}che}
\affiliation{SLAC National Accelerator Laboratory, Menlo Park, CA 94025, USA}

\author{Frank~Krauss}
\affiliation{Institute for Particle Physics Phenomenology, Durham University, Durham DH1 3LE, UK}

\author{Philipp Maierh{\"o}fer}
\affiliation{Physik--Institut, Universit{\"a}t Z{\"u}rich, CH--8057 Z{\"u}rich, Switzerland}

\author{Stefano Pozzorini}
\affiliation{Physik--Institut, Universit{\"a}t Z{\"u}rich, CH--8057 Z{\"u}rich, Switzerland}

\author{Marek~Sch{\"o}nherr}
\affiliation{Institute for Particle Physics Phenomenology, Durham University, Durham DH1 3LE, UK}

\author{Frank~Siegert}
\affiliation{Institut f{\"u}r Kern- und Teilchenphysik, TU Dresden, D--01062 Dresden, Germany}

\begin{abstract}
  We present differential cross sections for the production of top-quark pairs 
  in conjunction with up to two jets, computed at next-to leading order
  in perturbative QCD and consistently merged with a parton shower in the 
  \SherpaOpenLoops framework.  Top quark decays including spin correlation
  effects are taken into account at leading order accuracy.  The calculation 
  yields a unified description of top-pair plus multi-jet production, and 
  detailed results are presented for various key observables at the Large 
  Hadron Collider.  A large improvement 
with respect to the multi-jet merging approach at leading order
is found for the total transverse 
  energy spectrum, which plays a prominent role in searches for physics 
  beyond the Standard Model.
\end{abstract}

\maketitle

The top quark as the heaviest particle in the Standard Model is believed 
to play a fundamental role in many new physics scenarios. In a large variety 
of measurements at the Large Hadron Collider (LHC), top-quark events form 
either part of the signal or contribute a significant background in Higgs 
boson studies and new physics searches.  Top quarks are produced in abundance 
at the LHC, either in pairs or singly, and frequently 
in conjunction with several hard QCD jets.  Some first measurements of both 
inclusive production cross sections and of important kinematic distributions
have already been reported by the ATLAS and CMS 
experiments~\cite{Aad:2012hg,*Chatrchyan:2013faa}.
Top-quark pair production at hadron colliders suffers from large theoretical 
uncertainties at the leading order (LO) in perturbative QCD.
These uncertainties grow rapidly with the number of additional jets 
and represent a serious limitation for searches based on multi-jet signatures.
A number of precision calculations were completed recently, aimed at reducing 
these uncertainties: The inclusive production cross section
has been determined at next--\-to--\-next--\-to leading order (NNLO)
in the perturbative expansion~\cite{Czakon:2013goa}.
Parton-level predictions of top-quark pair production in association
with up to two jets have been computed at the next--\-to leading order (NLO)
in the strong coupling~\cite{Dittmaier:2007wz,*Bredenstein:2009aj,*Bredenstein:2010rs,*Bevilacqua:2009zn,*Bevilacqua:2010ve,*Bevilacqua:2011aa}, and 
NLO calculations for top-quark pair production in association with 
one jet~\cite{Kardos:2011qa,*Alioli:2011as} and with a
bottom-quark pair~\cite{Kardos:2013vxa,*Cascioli:2013era}
were matched to parton showers. 

The need for increasingly accurate and realistic simulations of $\ttbar+$jets 
production calls for a combination of parton showering with NLO calculations 
up to the highest possible jet multiplicity.  Addressing this need in this 
letter NLO matrix elements for the production of top-quark pairs in 
association with up to two jets are matched to the parton shower. 
Additionally, we also merge, for the first time, NLO matrix elements with
lower jet multiplicities, {\it i.e.}\ we combine $\ttbar$, $\ttbar j$ and
$\ttbar jj$, thereby extending previous results for $\ttbar+0,1$ 
jets~\cite{Frederix:2012ps,Hoeche:2013mua}.  This provides a fully inclusive simulation, which 
simultaneously describes $\ttbar+0$, $1$, $2$ jet configurations at NLO 
accuracy supplemented by the resummation of large logarithmic corrections 
provided by the parton shower.

Parton shower simulations in conjunction with LO QCD calculations of the hard 
scattering process have been the de-facto standard for computing observables 
at hadron colliders for decades.  Parton showers dress hard-scattering events 
with multiple emissions of QCD partons, thereby resumming large logarithmic 
corrections to all orders in perturbation theory.  Being based on 
the collinear approximation, they lack however a proper description of jet 
production at high transverse momenta or at wide angular separation.  The 
first techniques to remedy this deficiency were LO merging algorithms~\cite{
  Catani:2001cc,*Lonnblad:2001iq,*Krauss:2002up,*Mangano:2001xp,
  *Alwall:2007fs,*Hamilton:2009ne,Hoeche:2009rj}, which consistently combine
a description of multiple hard-jet emissions through higher-order 
tree-level matrix elements with the resummation of large soft and collinear
logarithms through the parton-shower.  Another method to improve 
parton-shower simulations consists of matching them to a full NLO 
calculation for a given final state~\cite{Frixione:2002ik,*Frixione:2003ei,
  Nason:2004rx,*Frixione:2007vw},
which yields NLO accurate predictions for observables that are inclusive with 
respect to extra jet radiation.  This method is however limited to improvements
to first order in the strong coupling and therefore does not lead to an improved 
description of multiple jet production, unless it is supplemented with a
suitable scale choice and Sudakov reweighting~\cite{Hamilton:2012np}.

Recent theoretical developments have led to new methods that combine the
complementary advantages of matching and merging, resulting in an NLO accurate 
description of final states with varying jet multiplicity~\cite{Hoeche:2012yf,
  *Gehrmann:2012yg,Lonnblad:2012ix,Frederix:2012ps}. One of these new NLO 
merging techniques, the \MEPSatNLO method~\cite{Hoeche:2012yf,*Gehrmann:2012yg}
is used in this publication. In this approach, NLO-matched simulations with 
increasing jet multiplicity are merged by vetoing emissions above a predefined 
hardness threshold, $Q_{\rm cut}$, denoted as merging scale. In analogy to LO 
merging, an optimal renormalization scale choice in presence of multiple jet 
emissions is defined, and the calculations with $n$ hard, well separated jets 
are made exclusive by means of appropriate Sudakov form factors.  
The $\mc{O}(\alpha_s)$ corrections generated by this procedure are consistently
subtracted in order to preserve both the fixed-order accuracy of the NLO 
calculations and the logarithmic accuracy of the parton shower~\cite{
  Lavesson:2008ah,Hoeche:2012yf,*Gehrmann:2012yg,Lonnblad:2012ix}.
This is a key improvement
that cannot be obtained through separate \SMCatNLO simulations based on 
$t\bar t$, $t\bar t j$ and $t\bar t jj$ matrix elements.
The parton-shower matching used in \MEPSatNLO presents a modified version of 
the original \MCatNLO algorithm~\cite{Frixione:2002ik}, called \SMCatNLO.  
It is based on including the fully coherent soft radiation pattern for the 
first emission~\cite{Hoeche:2011fd,*Hoeche:2012ft,*Hoeche:2012fm} by 
exponentiating dipole subtraction terms originally constructed for NLO 
calculations~\cite{Catani:1996vz,*Catani:2002hc}.   This is achieved through
a reweighting technique, which allows the generation of non-probabilistic
expressions as part of a Markov chain. 

The \MEPSatNLO simulation of $\ttbar+0,1,2$ jets presented in this letter 
merges multi-jet matrix elements at an unprecedented level of complexity. 
This is achieved by combining the event generator 
\Sherpa~\cite{Gleisberg:2008ta} with  \OpenLoops~\cite{Cascioli:2011va}, a 
fully automated one-loop generator based on a numerical recursion that allows 
the fast evaluation of scattering amplitudes with many external particles. 
For the numerically stable determination of both scalar and tensor integrals 
the \Collier library~\cite{collier} is employed, which implements 
the methods of~\cite{Denner:2002ii,*Denner:2005nn,*Denner:2010tr}.
The parton shower in \Sherpa is based on Catani-Seymour 
subtraction~\cite{Schumann:2007mg}.  The infrared subtraction
is performed by the dipole method~\cite{Catani:1996vz,*Catani:2002hc}
automated in both the \Amegic and \Comix modules of
\Sherpa~\cite{Gleisberg:2007md,Gleisberg:2008fv}, which also compute the
tree-level amplitudes and evaluate the phase-space integrals.
Top-quark decays are treated at LO including spin correlations 
based on $\ttbar$+jets Born matrix elements using spin density matrices 
\cite{Richardson:2001df,Hoche:2014kca}. Their kinematics are adjusted 
a posteriori according to a Breit-Wigner distribution using the top quark 
width as an input.

We simulate $\ttbar$+jets production at the 7 TeV LHC to be applicable to 
ongoing analyses. We use the MSTW 2008 
NLO PDF set~\cite{Martin:2009iq} and the corresponding strong coupling. 
Matrix elements are computed with massless $b$-quarks, but $b$-quark mass 
effects are consistently included in the parton shower.  According to the CKKW 
prescription~\cite{Hoeche:2009rj}, the renormalization scale for $\ttbar+n$ 
jet contributions is defined to be the solution of 
$\alpha_s(\mu_{\mathrm{R}})^{2+n}=\alpha_s(\mu_\text{core})^2\prod\alpha_s(t_i)$,
where the $\alpha_s$ terms associated with jet emissions are evaluated at the 
corresponding clustering scales $t_i$, while the scale associated with the 
$\pp\to\ttbar$ core process is defined by 
$1/\mu^2_\text{core}={1}/{s}+{1}/{(m_t^2-t)}+{1}/{(m_t^2-u)}$. 
$\mu_\text{core}$ is also used as factorization scale ($\mu_{\mathrm{F}}$) and as 
the parton-shower starting scale, $\mu_{Q}$. The merging scale is set to 
$Q_{\rm cut}=30$~GeV.  To assess theoretical uncertainties we rescale
$\mu_{\mathrm{R}}$ and $\mu_{\mathrm{F}}$ by factors of two, while $\mu_{{Q}}$ 
is varied by $\sqrt{2}$ and $Q_{\rm cut}$ is varied between 20 and 40~GeV. 
Additionally, intrinsic parton shower uncertainties are assessed by switching
between the two recoil schemes detailed in~\cite{Schumann:2007mg,Hoeche:2009xc}.
The combined renormalization- and factorization-scale uncertainty is
added in quadrature with 
the
other variations to form the total theoretical 
uncertainty.  Our results do not include the simulation of multiple parton
scattering or hadronization.  The publicly available version 2.1.0 of the 
\Sherpa event generator is used, and analyses are performed with 
\Rivet~\cite{Buckley:2010ar}.
The \OpenLoops program is publicly available at~\cite{hepforge}.

\begin{figure*}[t]
  \vspace*{-2mm}
  \subfigure[]{
    \begin{minipage}{0.4\textwidth}
      \lineskip-1.7pt
      \includegraphics[width=\textwidth]{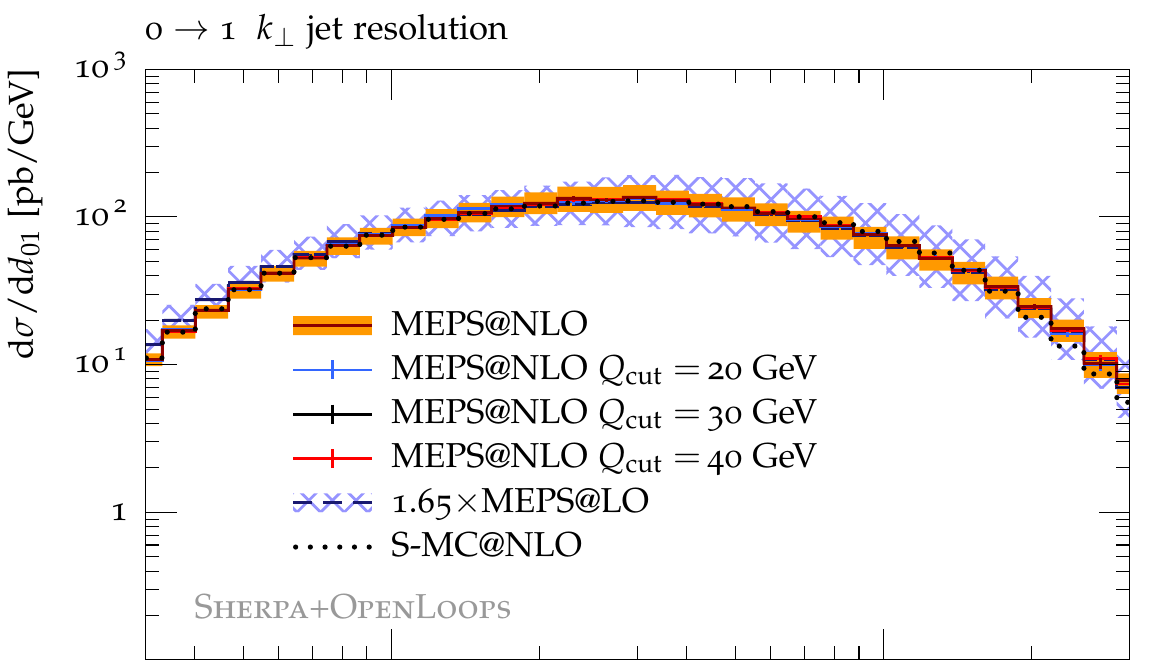}
      \includegraphics[width=\textwidth]{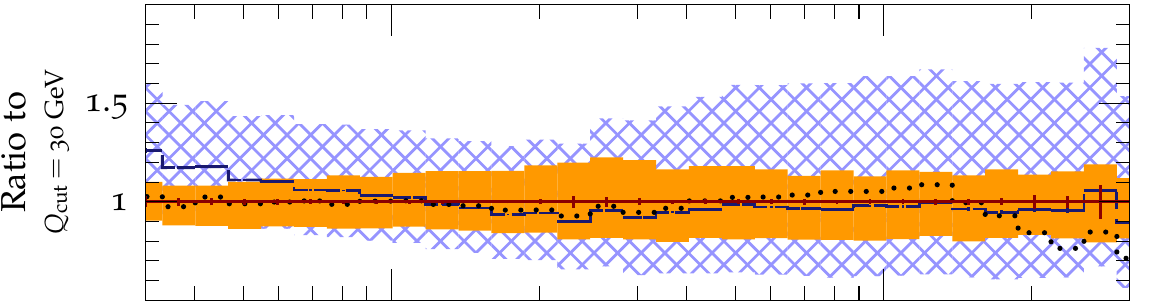}
      \includegraphics[width=\textwidth]{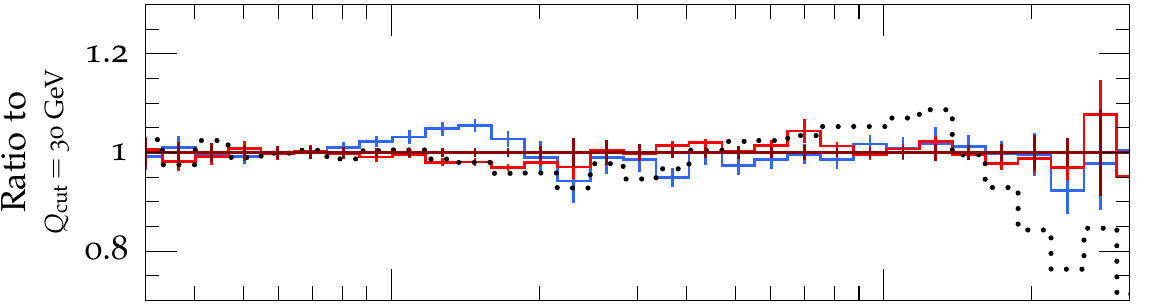}
      \includegraphics[width=\textwidth]{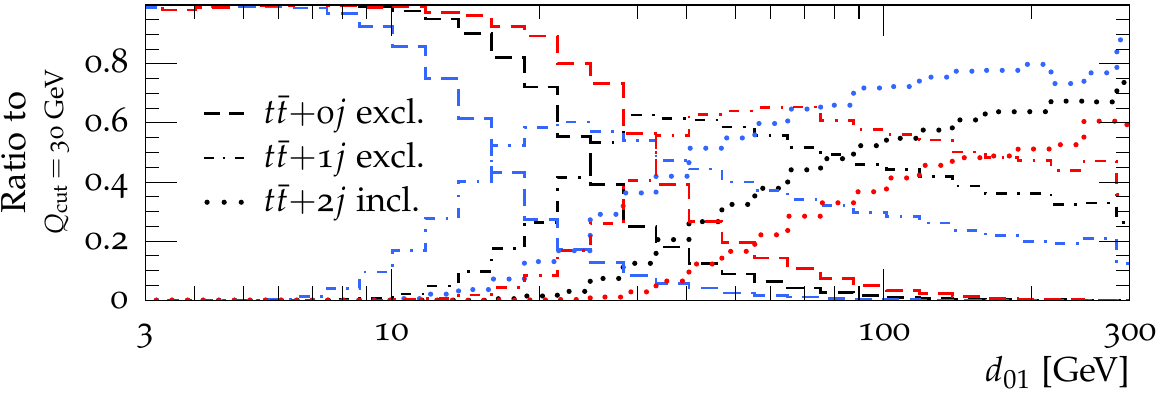}
    \end{minipage}
    \label{fig:kt-d01}}\qquad\qquad
  \subfigure[]{
    \begin{minipage}{0.4\textwidth}
      \lineskip-1.7pt
      \includegraphics[width=\textwidth]{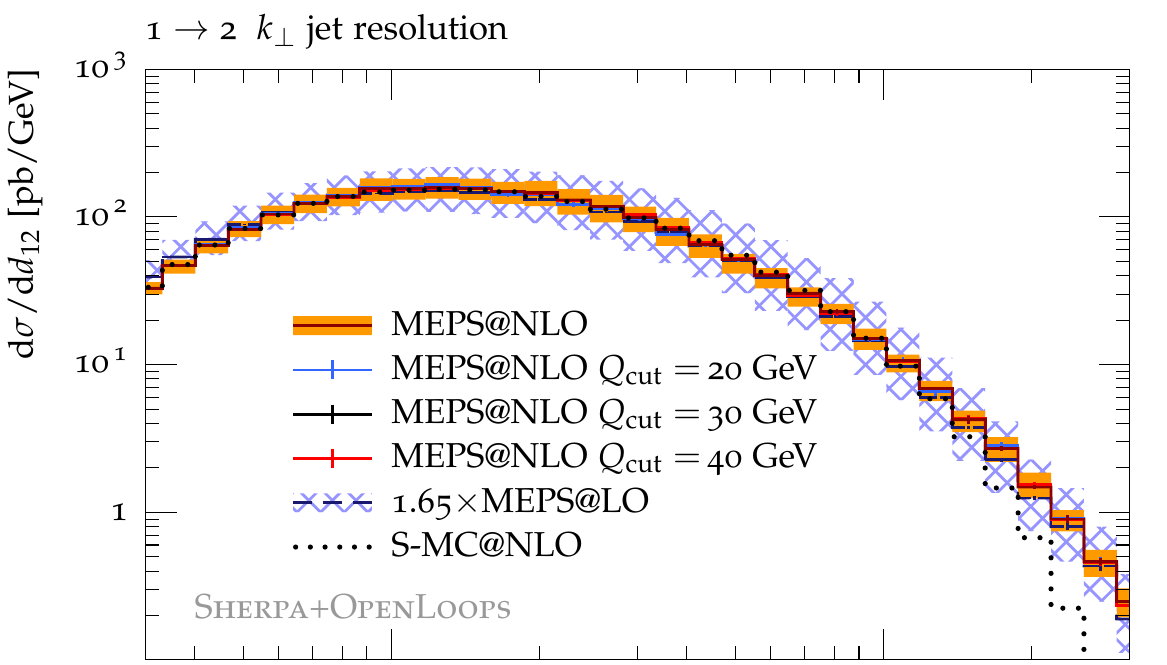}
      \includegraphics[width=\textwidth]{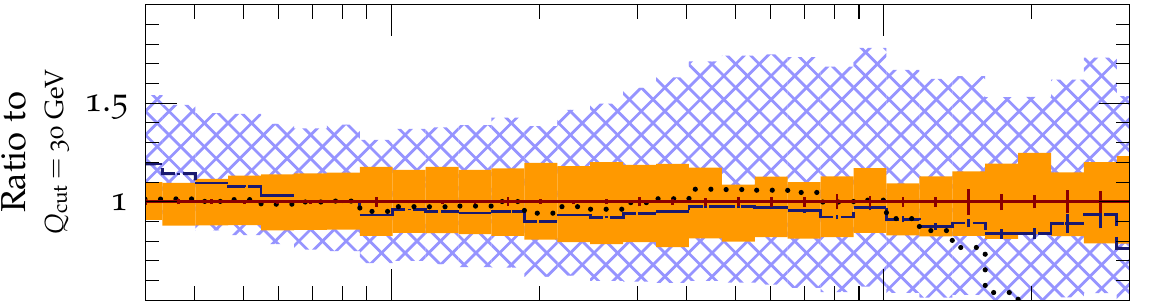}
      \includegraphics[width=\textwidth]{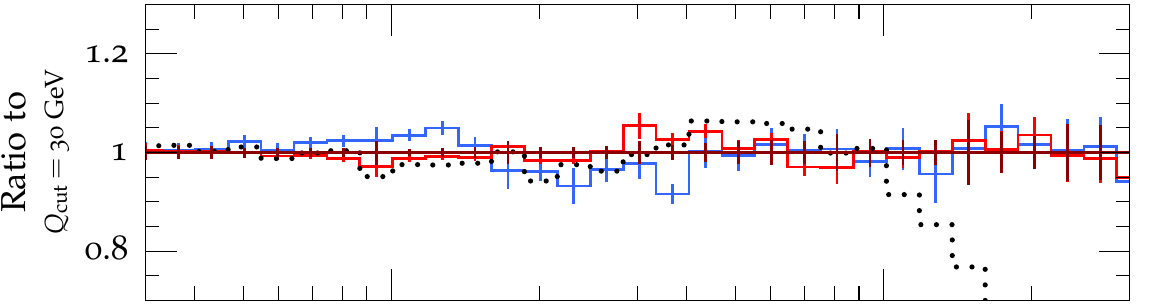}
      \includegraphics[width=\textwidth]{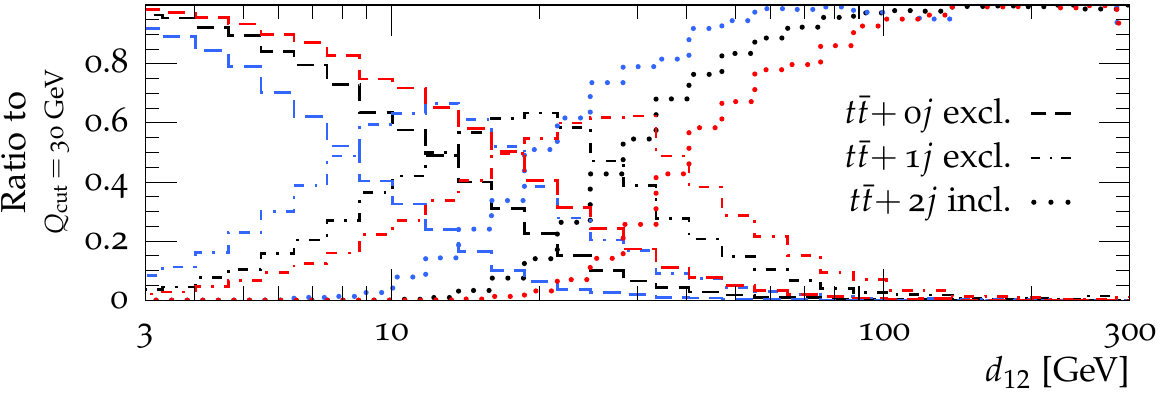}
    \end{minipage}
    \label{fig:kt-d23}}  \vspace*{-3mm}
  \caption{Differential $k_\perp$ jet resolutions calculated from all partons 
    (excluding top quarks) without any restrictions on their phase-space 
    in a calculation with stable top final states. Solid 
    lines indicate the \protect\MEPSatNLO prediction for three different 
    values of the merging scale, $Q_\text{cut}=$ 20 GeV (blue), 30 GeV (red), 
    and 40 GeV (green). They are contrasted with the combined 
    $\mu_R$-$\mu_F$-$\mu_Q$-uncertainties of the central \protect\MEPSatNLO 
    (orange full band) and \protect\MEPS (blue hatched band) predictions. The 
    center ratio highlights the $Q_\text{cut}$-variation only,
    while the lower ratio details the relative contributions of the individual 
    matched $pp\to t\bar{t}+n\,\text{jets}$ calculations and how they vary with $Q_\text{cut}$.
    Statistical uncertainties are indicated by error bars.\label{fig:fig1}}
\end{figure*}

\begin{table}[t!]
  \begin{tabular}{l|c}
    Method & $\sigma_\text{incl}$ \\\hline
    \MEPSatNLO \Hl& \resseperrs{1.85}{0.31}{0.04}{0.03}{0.00}{0.30}{0.05}{0.01}{0.00}\\
    \MEPS      \Hl& \resseperrs{1.11}{0.32}{0.04}{0.03}{0.00}{0.55}{0.05}{0.01}{0.02}\\
    \SMCatNLO  \Hl& \resseperrs{1.85}{0.23}{0.00}{0.00}{0.00}{0.26}{0.00}{0.00}{0.00}
  \end{tabular}
  \caption{Inclusive cross section and its uncertainties originating from a
    variation of $\mu_R$ and $\mu_F$, $\mu_Q$, $Q_\text{cut}$ and the parton 
    shower recoil scheme, in that order, as detailed in the 
    text.\label{tab:tab1}}
\end{table}

To evaluate the quality of our multi-jet merged calculation the inclusive 
cross section and the size of its uncertainties stemming from the afore 
mentioned four 
different sources is examined in Tab. \ref{tab:tab1}. As can 
be seen, the \MEPSatNLO calculation reproduces the inclusive cross section 
of the \SMCatNLO calculation well. In each case, the uncertainties are 
dominated by the renormalization and factorization scale variations while 
the dependence on merging parameter $Q_\text{cut}$ is minimal. 
This demonstrates that using a relatively small 
merging scale does not lead to a significant loss of accuracy 
within the \MEPSatNLO framework.
In this respect, let us remind that 
inclusive cross sections involve uncontrolled logarithms of the merging scale 
that are subleading with respect to the
aimed NLO+NLL accuracy, but can become of 
order $\sqrt{\alpha_S}$ when the merging scale is as small as the
location of the Sudakov peak~\cite{Hamilton:2012rf},
which is around 10~GeV in top-pair production. 
This observation, which is referred to as formal loss of NLO accuracy 
in~\cite{Hamilton:2012rf}, relies on an argument 
of purely formal nature, in the sense that the ``uncontrolled'' logarithms
involve an unknown coefficient, which is generically assumed to be of order
one.  In practice, the fact that we observe a  $Q_\text{cut}$ dependence at the 
few percent level,\footnote{
Let us note that 
a standard method to quantify
merging scale uncertainties does not exists to date,
and our assessment depends on the choice of the variation range
20~GeV$< Q_\text{cut}<$ 40~GeV.
A possible systematic approach to assess merging uncertainties 
for the case of small merging scales was sketched in~\cite{nnlomc}.
When applied to \MEPSatNLO for $t\bar t$+jets, this method suggest
that  merging scale uncertainties remain below ten percent down to merging scales of
the order of 10~GeV.
}
i.e.~well below the NLO uncertainty associated with renormalisation 
and factorization  scale variations,
 suggests that the above mentioned coefficient is 
rather small in the case of \MEPSatNLO merging for $t\bar t+$jets.

To further investigate the properties of the multi-jet merging 
Fig.\ \ref{fig:fig1} contrasts 
the variation of $Q_\text{cut}$ with the combined uncertainty of the other 
three sources of uncertainties in a calculation where (only for this purpose) 
the final state tops are considered stable. As their missing decay kinematics 
do not introduce further jet activity the two observables found to be most sensitive 
to merging effects are the differential $0\to 1$ and $1\to 2$ $k_\perp$ jet 
resolutions ($R=0.6$), defined on all final state QCD partons (except the 
stable top quarks). Short-comings of the 
merging would show up as kinks at $d_{i(i+1)}\sim Q_\text{cut}$. Again, the 
dependence on $Q_\text{cut}$ is found to be smaller than the dependence 
on the other three 
sources of uncertainties combined. 
The latter does not exceed the 20\% level in the whole $k_T$ spectrum, both for the 
first and for the second jet.

\begin{figure*}[p]
  \vspace*{-2mm}
  \subfigure[]{
    \begin{minipage}{0.4\textwidth}
      \lineskip-1.7pt
      \includegraphics[width=\textwidth]{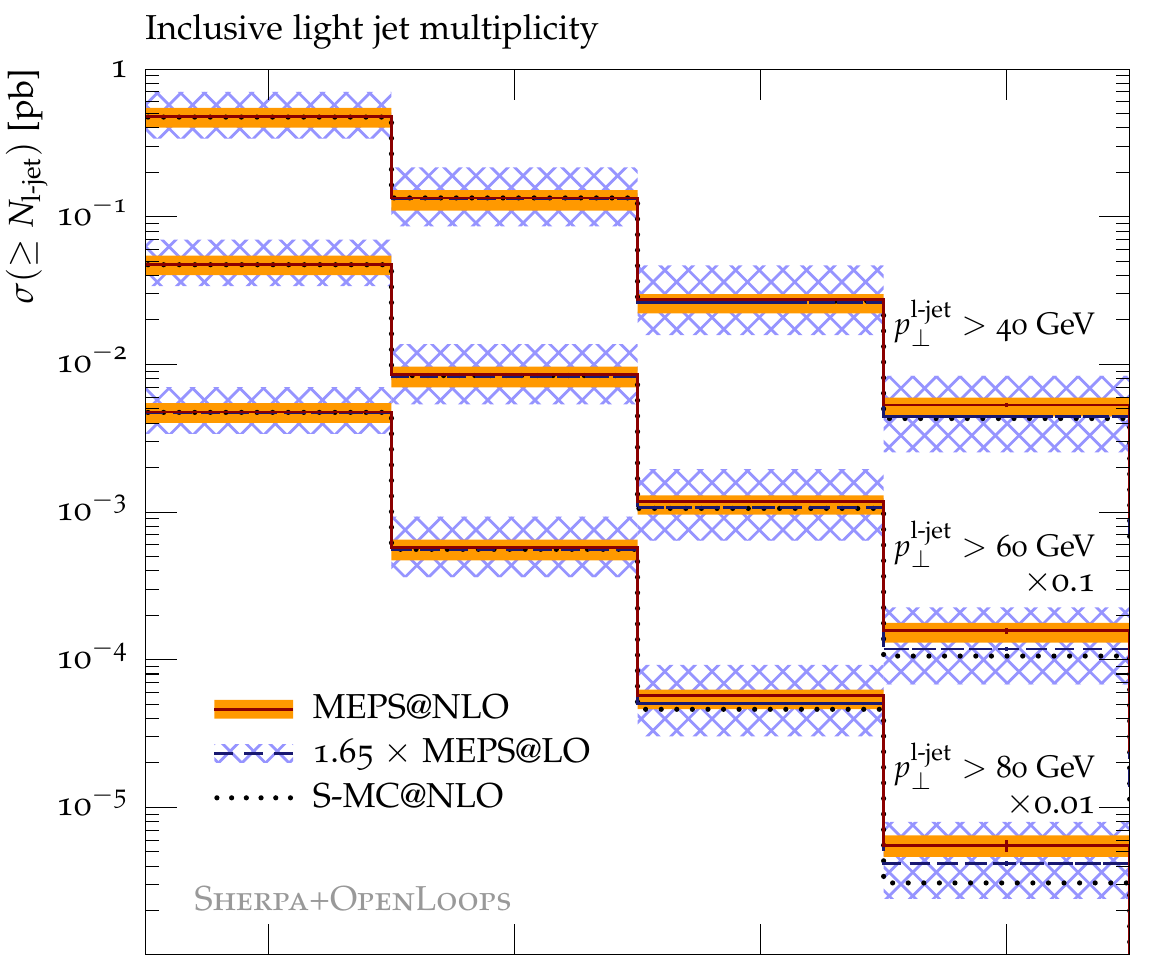}
      \includegraphics[width=\textwidth]{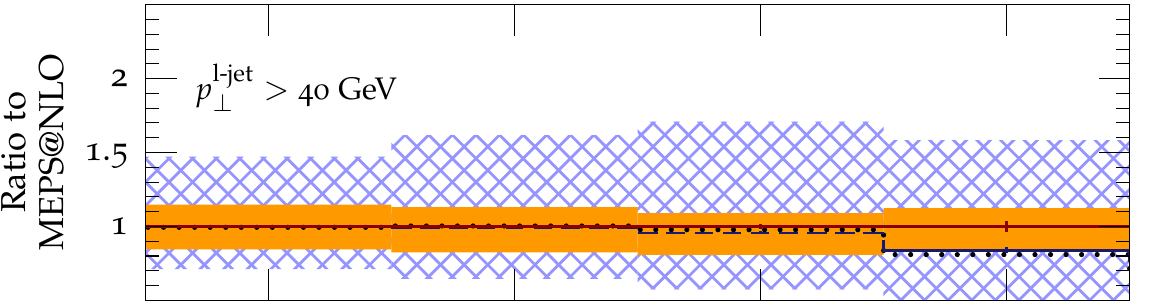}
      \includegraphics[width=\textwidth]{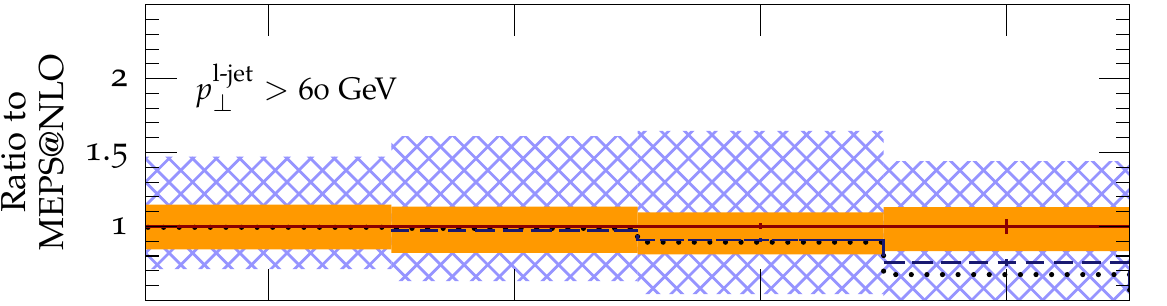}
      \includegraphics[width=\textwidth]{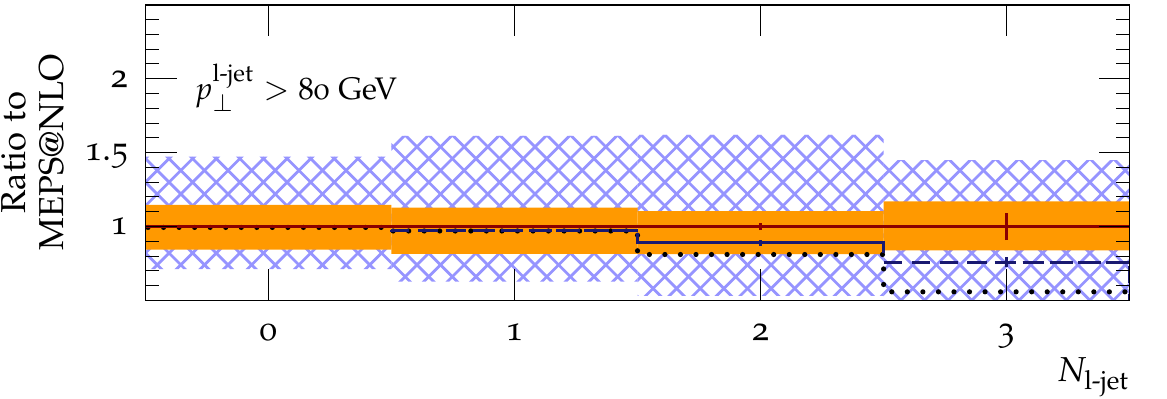}
    \end{minipage}
    \label{fig:jet_multi}}\qquad\qquad
  \subfigure[]{
    \begin{minipage}{0.4\textwidth}
      \lineskip-1.7pt
      \includegraphics[width=\textwidth]{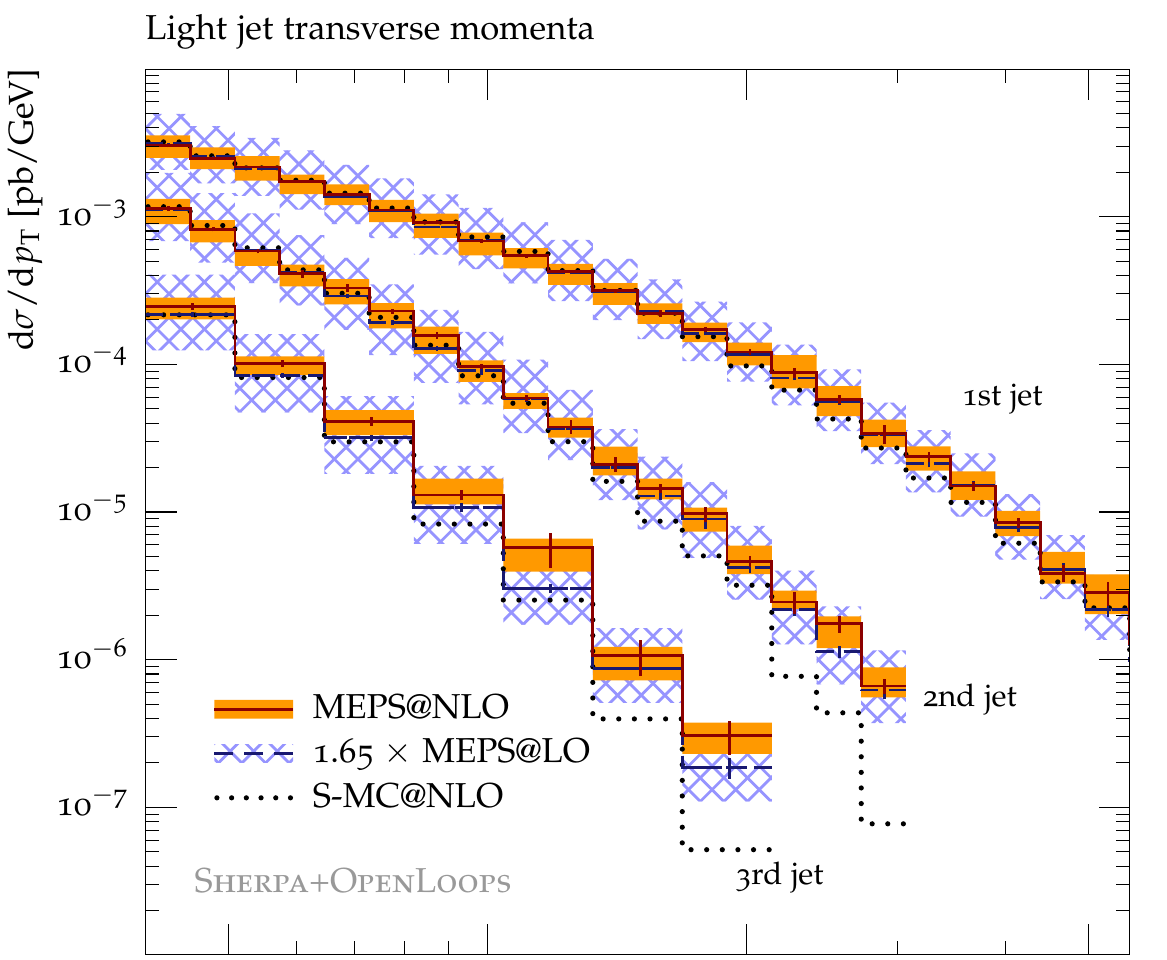}
      \includegraphics[width=\textwidth]{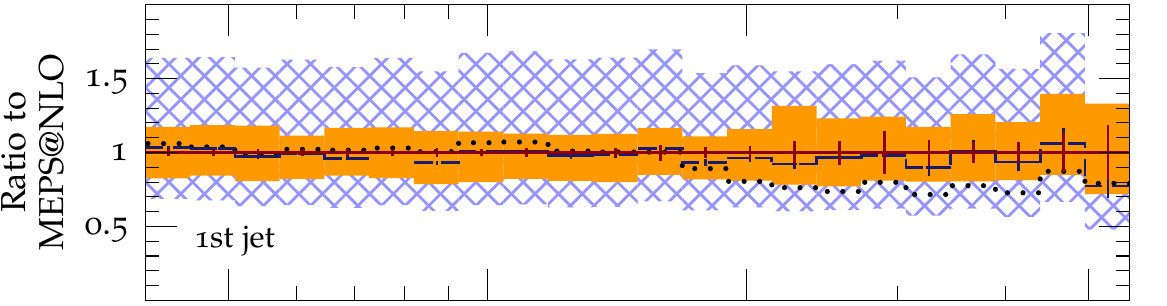}
      \includegraphics[width=\textwidth]{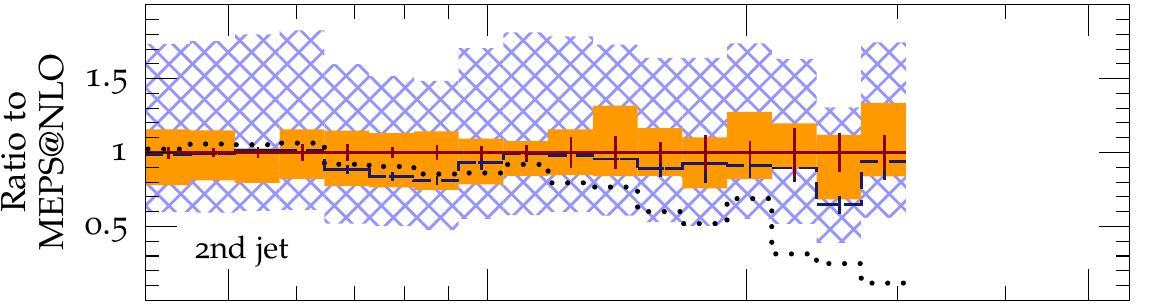}
      \includegraphics[width=\textwidth]{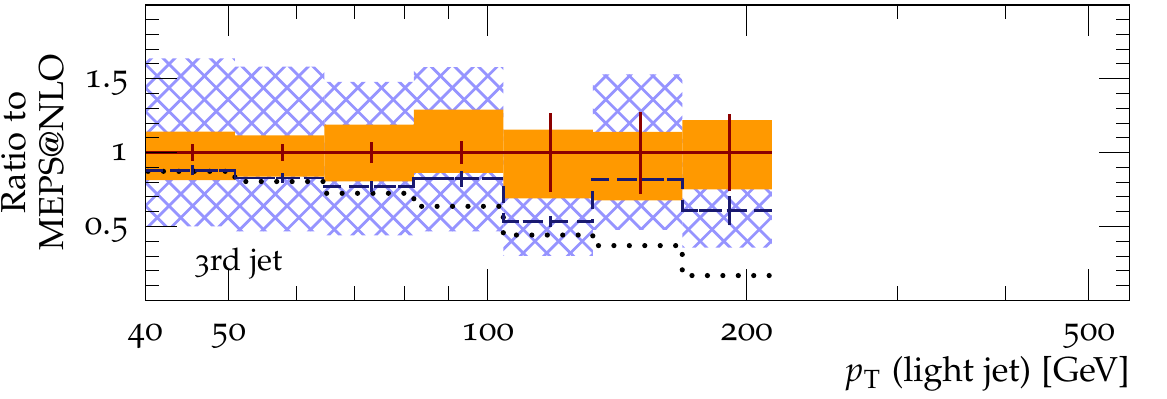}
    \end{minipage}
    \label{fig:jet_pt}}
  \vspace*{-3mm}
  \caption{Light-flavor jet multiplicity distribution (including $c$- but not 
    $b$-jets) for transverse momentum thresholds of 40, 60 and 
    80~GeV~\subref{fig:jet_multi} and transverse momentum spectra of the 
    three leading light-flavor jets~\subref{fig:jet_pt}.  Solid (red) lines 
    indicate \protect\MEPSatNLO predictions, and the full (orange) band shows 
    the corresponding total theoretical uncertainty.  
    Dashed lines indicate \protect\MEPS predictions, with the corresponding
    uncertainties shown as hatched (blue) bands. \protect\SMCatNLO predictions
    are shown as dotted histograms. Statistical uncertainties for each 
    calculation are indicated by error bars.\label{fig:fig2}}
\end{figure*}

\begin{figure*}[p]
  \vspace*{-2mm}
  \subfigure[]{
    \begin{minipage}{0.4\textwidth}
      \lineskip-1.7pt
      \includegraphics[width=\textwidth]{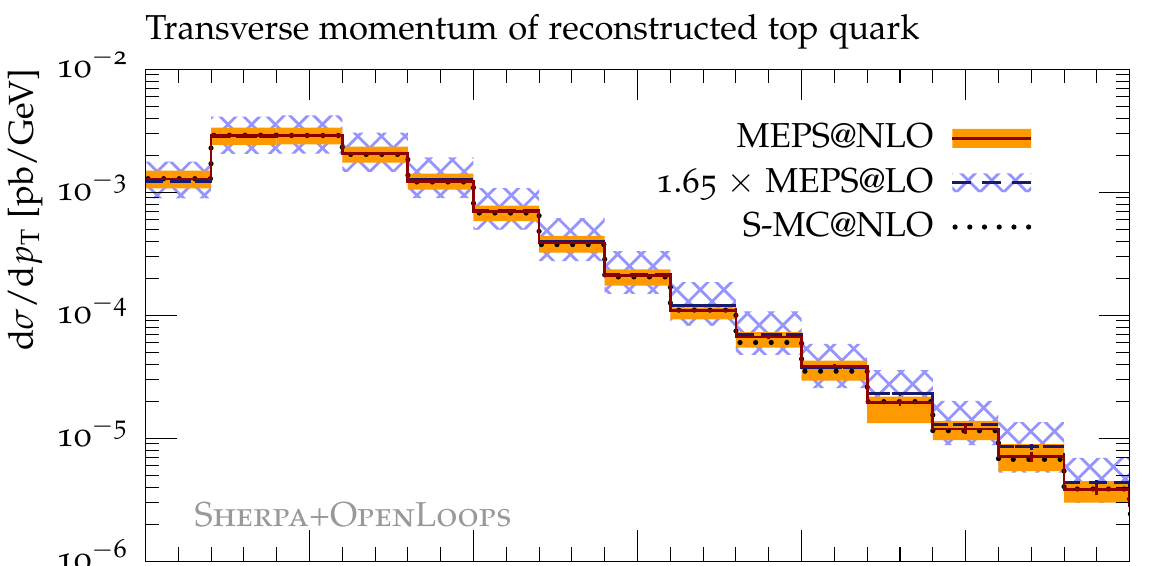}
      \includegraphics[width=\textwidth]{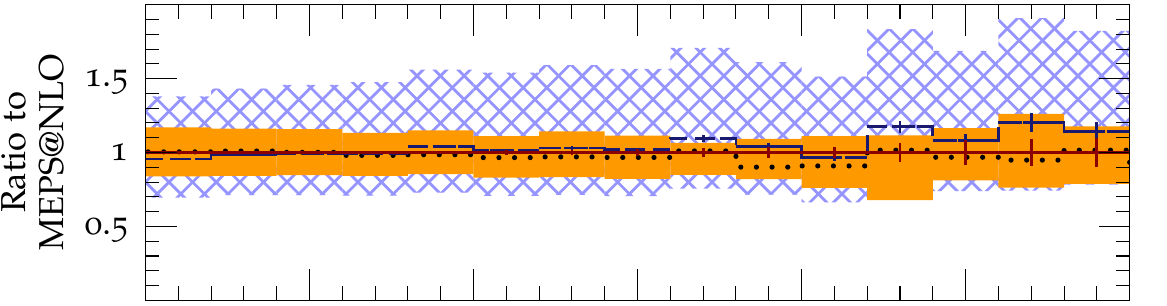}
      \includegraphics[width=\textwidth]{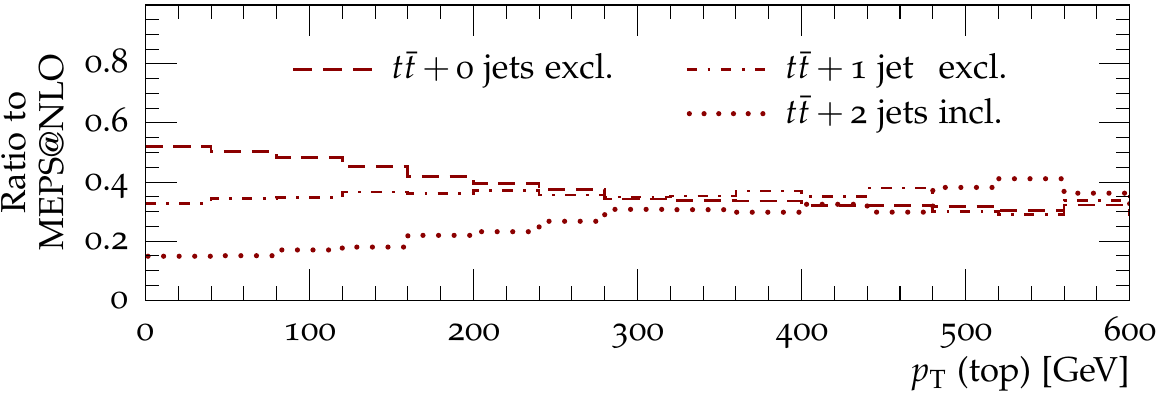}
    \end{minipage}
    \label{fig:top_pt}}\qquad\qquad
  \subfigure[]{
    \begin{minipage}{0.4\textwidth}
      \lineskip-1.7pt
      \includegraphics[width=\textwidth]{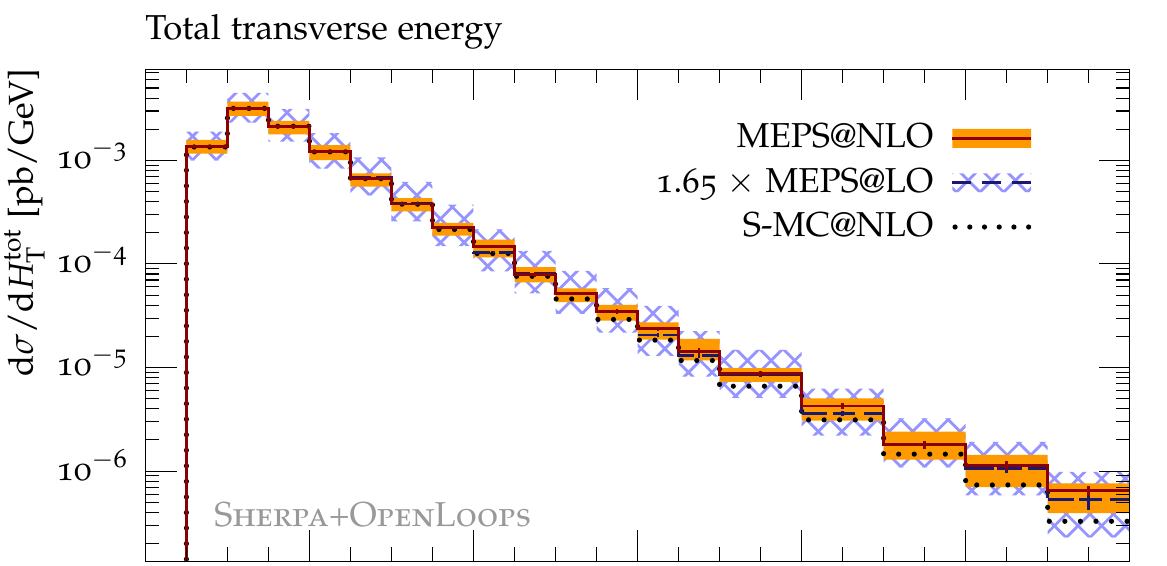}
      \includegraphics[width=\textwidth]{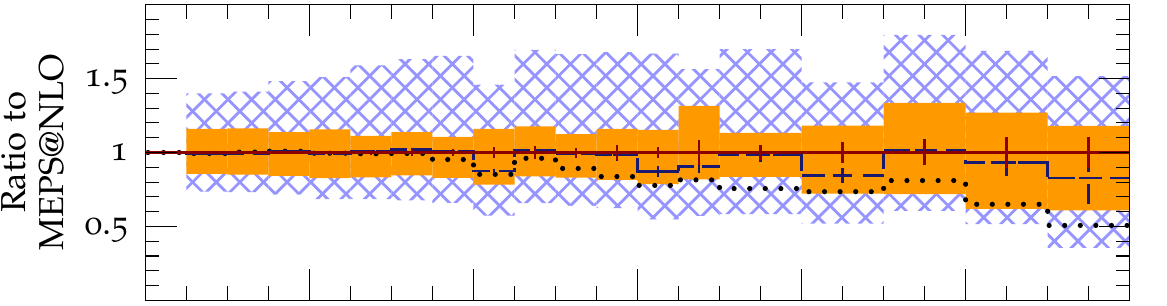}
      \includegraphics[width=\textwidth]{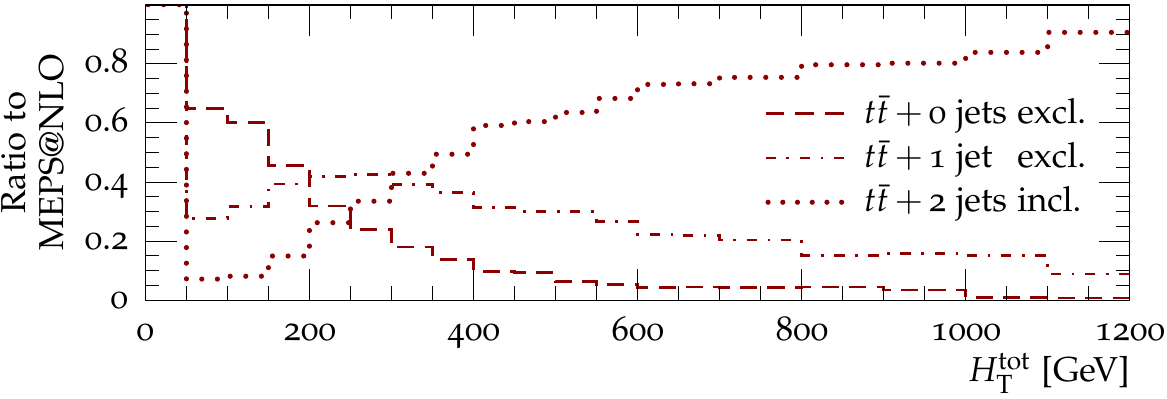}
    \end{minipage}
    \label{fig:ht}
  }
  \vspace*{-3mm}
  \caption{
    Transverse momentum of the reconstructed top quark~\subref{fig:top_pt} 
    and total transverse energy~\subref{fig:ht}, 
    see Fig.~\ref{fig:fig2} for details. The lower ratio details 
    the contributions of the individual 
    matched $pp\to t\bar{t}+n\,\text{jets}$ calculations.
    \label{fig:fig3}}
\end{figure*}

To perform a realistic analysis,
we identify the top quarks through their full decay final state and select 
events containing a positron and a muon with $p_{\rm T}>25$~GeV and $|\eta|<2.5$,
$E_\mathrm{T}^\text{miss}>30$~GeV is directly reconstructed from the neutrinos.  
Jets are defined using the anti-$k_t$ algorithm~\cite{Cacciari:2008gp} with 
$R=0.4$.  Ideal $b$-jet tagging is modeled based on the flavor of the jet 
constituent partons.  
Defining the sign of each $b$-jet according to its $b$-quark contents, 
exactly one $b$- and one anti-$b$-jet with $p_{\rm T}>25$~GeV and
$|\eta|<2.5$ are required. 

Figures~\ref{fig:fig2}-\ref{fig:fig3} feature various observables that characterize multiple light-jet 
emissions in this $\ttbar$+jets event selection.  Our best predictions, based 
on \MEPSatNLO next-to-leading order merging, are compared to leading-order 
merged results (\MEPS), evaluated in an identical setting but rescaled by 
the inclusive $K$-factor of 1.65, and to an inclusive \SMCatNLO simulation 
for $\pp\to \ttbar$.  The latter two simulations represent the typical level of
theoretical accuracy that is currently attained in the analysis of LHC 
data. It is important to point out that all differential observables discussed 
in the following are dominated by different exclusive jet multiplicity 
calculations in different regions, necessitating a multi-jet merged approach to 
achieve the highest accuracy throughout.
This is exemplified by detailing the composition of the obervables 
of Fig.\ \ref{fig:fig3} in terms of their input matched $pp\to t\bar{t}+n\;\text{jets}$ 
calculations in the lower ratio panel.

Uncertainty estimates are only shown for the merging approaches, while the
\SMCatNLO matching approach does not allow for a realistic uncertainty estimate
in multi-jet final states, whose description is entirely based on parton
shower emissions.\footnote{
  At present there is no parton shower implementation that allows for consistent 
  scale variations for the $\alpha_S$ terms associated with shower emissions. 
  Moreover, it is not possible to assess the uncertainty related to the 
  underlying soft and collinear approximations in the parton shower framework.
}
In fact, the systematic reduction of theory uncertainties
and the possibility to estimate them in a realistic way through scale
variations is one of the main advantages of NLO multi-jet merging.

The multiplicity distribution of light-flavor jets is displayed for 
thresholds of $p_{\rm T}>40$, $60$ and $80$~GeV in Fig.~\ref{fig:jet_multi}.  
As compared to \MEPS, the uncertainty of the inclusive \MEPSatNLO
cross section within acceptance cuts is steeply reduced from 48\% to 16\%,
while that for events with at least one light-flavor jet of 
$p_{\rm T}>40/60/80$~GeV is reduced from 64/65/66\% to 18/18/18\%.
Particularly
striking is the reduction in the uncertainty of the cross section of
producing the $\ttbar$-pair in association with at least two jets:
79/81/82\% to 19/19/19\%.
The $Q_{\rm cut}$ dependence of \MEPSatNLO predictions is typically 
well below ten percent, while the combined theoretical uncertainty 
is dominated by renormalization scale variations.

The jet transverse momentum distributions are shown in
Fig.~\ref{fig:jet_pt}.  
For the first two jets, \MEPSatNLO predictions feature 
scale variations of about 20\%. 
Apart from a slight increase in the hard region, which is in part due
to statistical fluctuations,
these uncertainties are rather independent of $p_T$.
For the third jet the uncertainty tends to be 
similarly small as for the first two ones, 
especially at low transverse momenta.
This is due to the fact that at relatively soft transverse momenta
the production of the third jet 
proceeds predominantly via
parton shower emissions on top of
$t\bar tjj$ events, which are described in terms of
NLO accurate matrix elements. 
Let us note that the resulting uncertainty 
is not fully realistic since, as pointed out above,
uncertainties associated with parton shower emissions are not 
correctly reflected by the standard scale variation approach.

Figure~\ref{fig:top_pt} shows the transverse momentum of the reconstructed
top quark. Again we observe a strong reduction of uncertainties, 
particularly at larger transverse momenta. This will significantly increase 
the precision in measurements of Standard Model $\ttbar$ production.
Finally, we analyze the total transverse energy, 
$H_{\rm T}^\text{tot}=\sum p_{\rm T,\text{b-jet}}
             +\sum p_{\rm T,\text{l-jet}}
             +\sum p_{\rm T,\text{lep}}
             +E_{\rm T}^\text{miss}$,
of the full final state, where only light jets with $p_T>40$~GeV are taken into
account. 
This observable plays a key role in searches for new physics, and 
its high sensitivity to QCD radiation requires accurate modeling of
multi-jet emissions. Fig.~\ref{fig:ht} shows a strong reduction of 
perturbative uncertainties, especially in the high-$H_{\rm T}^\text{tot}$ region.
We believe that this makes \MEPSatNLO the prime tool for 
estimating the theoretical precision of multi-jet merged
predictions in $\ttbar+$jets backgrounds to new-physics searches.
It is worth mentioning that for various observables in Figs.~1--2 the 
\MEPSatNLO, \MEPS and \SMCatNLO predictions agree remarkably 
well. This encourages the use of the less compute-intensive 
LO merging technique for making nominal predictions including 
full detector simulation in experimental analyses.
Purely matched calculations, such as \SMCatNLO are generally insufficient. 
This can be seen, for example, in the $\pT$-distributions, 
which are systematically low in the case of the 2nd and 3rd jet.

In summary we have presented the first unified simulation of top-quark pair
production in association with up to two jets including top-quark decays and
merging with the parton shower at the next-to-leading order in perturbative
QCD.  Residual theoretical uncertainties are reduced 
to the level of 
20\%.
A wide range of experimental analyses based on
multi-jet final states can strongly benefit from this improvement.
In particular, as compared to simulations based on 
multi-jet merging at leading order,
we observe a drastic reduction of
uncertainties for large values of the total transverse energy, $H_{\rm
T}^\text{tot}$, which is highly relevant for new physics searches at
the Large Hadron Collider.

\smallskip

\begin{acknowledgments}
We are grateful to A.~Denner, S.~Dittmaier and L.~Hofer for providing 
us with the \Collier library.
This work was supported by the SNSF, by the US Department of Energy under contract 
DE--AC02--76SF00515, and by the European Commission through the networks
PITN--GA--2010--264564, PITN--GA--2012--315877, and
PITN--GA--2012--316704.  We used
resources of the Open Science
Grid, supported by the National
Science Foundation and the U.S.\ Department of Energy's Office of Science 
\cite{Bauerdick:2012xd}.
\end{acknowledgments}
\vspace*{-3mm}

\bibliography{journal}

\end{document}